\documentclass[vci-paper]{vciart}
\usepackage{amssymb}
\usepackage{graphics}

\begin{document}

\vspace*{-1.8cm}
\begin{flushright}
{\large\bf LAL 04-14}\\
{\large\bf DAPNIA 04-78}\\
\vspace*{0.1cm}
{\large May 2004}
\end{flushright}

\begin{frontmatter}

\title{ION BACKFLOW IN THE MICROMEGAS TPC FOR THE FUTURE LINEAR COLLIDER}
\author[A]{P. Colas}, 
\author[A]{I. Giomataris}, 
\author[B]{V. Lepeltier},

\address[A]{DAPNIA, CEA Saclay, 91191 Gif sur Yvette C\'edex, France}
\address[B]{LAL, IN2P3 and Universit\'e de Paris-Sud, 91898 Orsay C\'edex, France}

\begin{abstract}
{We present ion backflow measurements in a Micromegas (MICRO-MEsh GASeous detector) TPC device developed for the next high energy electron-positron linear collider under study and a simple explanation for this backflow. A Micromegas micro-mesh has the intrinsic property to naturally stop a large fraction of the secondary positive ions created in the avalanche. It is shown that under some workable conditions on the pitch of the mesh and on the gas mixture, the ion feedback is equal to the field ratio (ratio of the drift electric field to the amplification field). Measurements with an intense X-ray source are in good agreement with calculations and simulations. The conclusion is that in the electric field conditions foreseen for the Micromegas TPC (drift and amplification fields respectively equal to 150-200 V/cm and 50-80 kV/cm) the expected ion backflow will be of the order of $2-3 \times 10^{-3}$. In addition, measurements have been done in a 2T magnetic field: as expected the ion backflow is not altered by the magnetic field.}
\end{abstract}
\end{frontmatter}

\section{The Micromegas TPC}
In a TPC, especially in high background conditions, it is very important to have a very limited ion backflow from the secondary ions produced in the amplification region to the drift volume, in order to avoid distortions of the drift electric field. The MWPC TPC's (as ALEPH, DELPHI or STAR) are equipped with a gating grid, where two consecutive wires are polarised at opposite voltages, so creating a transverse field stopping most of the secondary ions before they reach the drift space.

\begin{figure}[h]
\centering{\resizebox{8.cm}{!}{\includegraphics{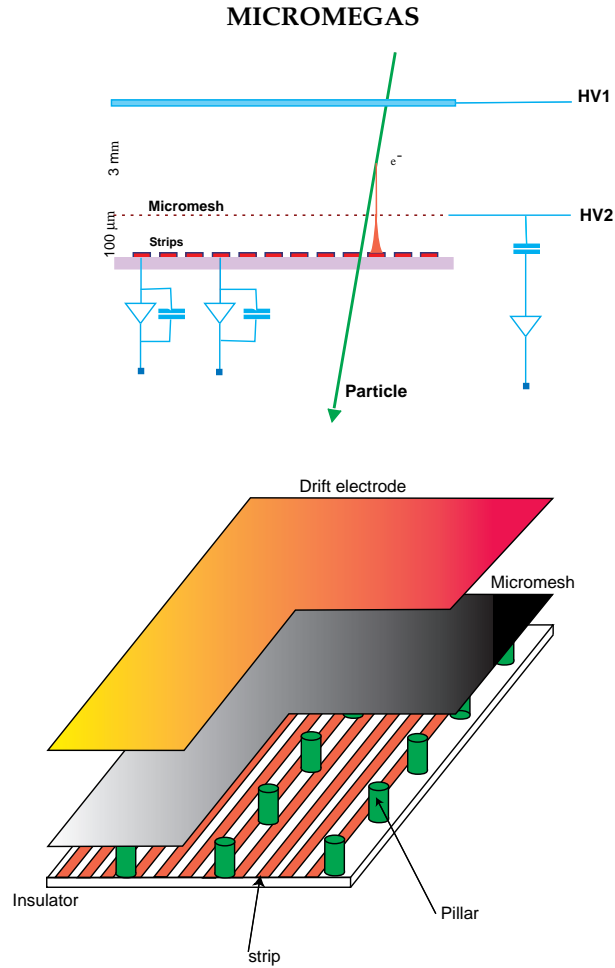}}}
\caption{Micromegas device}
\label{fig:Fig1}
\end{figure}

 For the physics to be studied on the next linear collider \cite{ref1}, it is proposed to build a high performance large TPC, using instead of MWPC, new MPGD (Micro-Pattern Gaseous Detector) readout. The Micromegas \cite{ref2} (MICRO-MEsh GASeous detector), under development for the future TPC \cite{ref3}\cite{ref4} by the Saclay-Orsay collaboration is a parallel plate device, simply composed of a very thin (5 $\mu$m) metallic micro-mesh, with a pitch of 25 to 50 $\mu$m, set at a small distance from the anode plane (50-100 $\mu$m). Primary electrons coming from the drift space cross the micromesh, which is fully transparent, and avalanche in the small gap, where a voltage of $\sim$300-500 V is applied between the two electrodes (see figure \ref{fig:Fig1}).

\vspace*{0.5cm}

\section{Theory and predictions}
\begin{figure}[h]
\centering{\resizebox{9.cm}{!}{\includegraphics{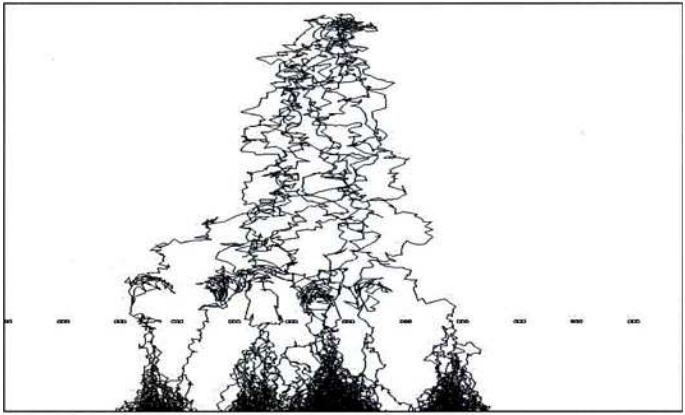}}}
\caption{GARFIELD simulation of electron drift and multiplication in Micromegas}
\label{fig:Fig2}
\end{figure}

\vspace{-5mm}  
Micromegas has many advantages: among them, a fast signal, no ExB effect, a high gain, a good energy resolution; it has also the capability to naturally stop most of the ions produced in the amplification space. Due to the very large field ratio $\alpha$ between the multiplication and the drift regions (as high as 400 or 500) the electric field lines are very much compressed between the two regions ("funnel" effect). Following the Gauss theorem, the compression factor of the field lines is equal to the field ratio $\alpha$. But due to collisions in the gas, electrons do not drift along the field lines. They diffuse, especially also in the multiplication space: the transverse extension $\sigma$ (standard deviation) of the avalanche due to diffusion is of the order of 10-15 $\mu$m, depending on the gas mixture, the electric field and the gap width; in figure  \ref{fig:Fig2}  is shown a simulation by GARFIELD of electron diffusion and multiplication in the drift and the multiplication gaps. 

\begin{figure}[h]
\centering{\resizebox{9.cm}{!}{\includegraphics{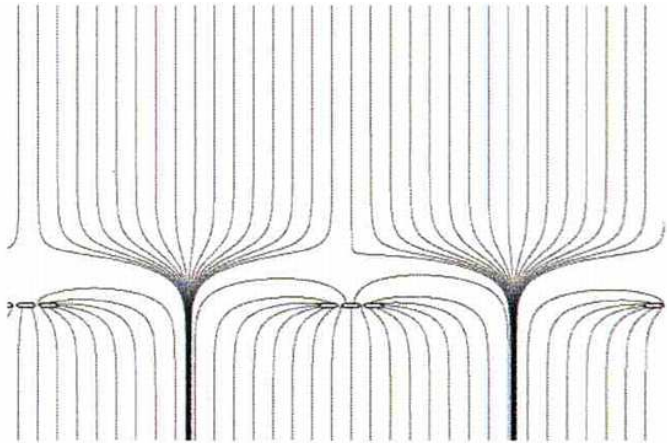}}}
\caption{Field lines in Micromegas}
\label{fig:Fig3}
\end{figure}

This electron cloud size is much larger than the size of the funnel end \linebreak (1-2 $\mu$m in radius in the TPC conditions).Conversely, ions, due to their high mass, are not submitted to diffusion and drift along the field lines. Assuming that they are emitted with the same distribution as the avalanche, most of them are naturally collected by the micro-mesh (see figure \ref{fig:Fig3}), and only the fraction of ions created inside the small funnel will flow back into the drift volume. 

\vspace*{0.4cm}

Analytic calculations have been done assuming a bi-dimensional Gaussian distribution of the electron diffusion in the multiplication space. It is assumed that ions are created from the anode plane, with the same Gaussian distribution (rms $\sigma$) as the avalanche. This is valid since the gain is generally large enough (at least a few hundred), and most of the ions are emitted at a very small distance (a few $\mu$m only) from the anode plane. Then ions are supposed to drift along the field lines without any diffusion. Ions created outside the funnel will follow field lines ending on the micro-mesh, and will be naturally collected by it; a very small fraction, produced inside the thin funnel, will drift along field lines flowing from the drift volume, and will feed it, before being collected by the HV electrode of the TPC after a very long time (typically a few hundred ms for a 2m drift length).
 
\begin{figure}[h]
\vspace*{0.3cm}

\centering{\resizebox{10.cm}{!}{\includegraphics{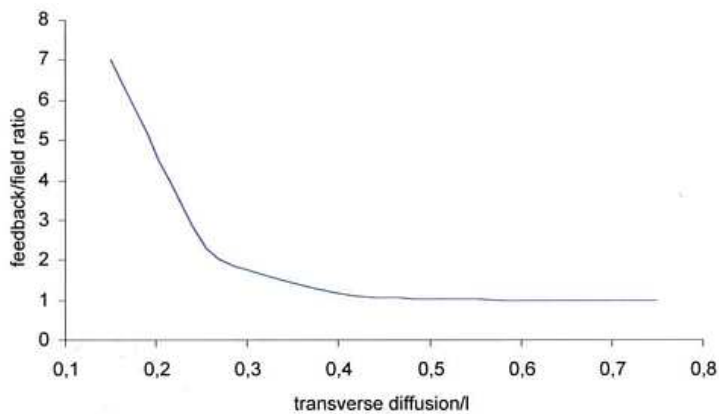}}}
\caption{Computed value of $\alpha$ (field ratio times ion feedback) as a function of $\sigma$/l (transverse diffusion divided by mesh pitch)}
\label{fig:Fig4}
\end{figure}

   Following the previous assumptions, it is easy to compute the ion backflow fraction $\beta$ as a function of the field ratio $\alpha$. As expected the key parameter is the relative value of the size of the ion cloud ($\sigma$) and the mesh pitch \cite{ref1}. On figure \ref{fig:Fig4} is shown the product $\beta\alpha$ as a function of $\sigma$/l: if this parameter is small (small diffusion and/or too small pitch mesh) ion feedback $\beta$ is substantially larger than the inverse of the field ratio $\alpha$; if $\sigma$/l is greater than 0.5, the optimal is reached, with an ion feed back equal to 1/$\alpha$.

\vspace*{0.3cm}

\vspace*{0.5cm}

   This condition is easily reached: for most usual gas mixtures, the transverse diffusion at high electric field (40-70 kV/cm) is of the order of 120-150 $\mu$m$\times$ cm$^{-1}$  ie $\sigma$ = 12-15 $\mu$m for a 100 $\mu$m amplification gap. With a 500 lpi (lines per inch) micromesh (50 $\mu$m pitch), $\sigma$/l is equal to .25-.3, and the ion backflow is 2 or 3 times larger than the optimal value 1/$\alpha$. With a 1000 or 1500 lpi mesh (25 or 17 $\mu$m pitch) $\sigma$ /l is larger than .5, and the expected feedback is equal to the inverse of the field ratio $\alpha$. 
  
   As a conclusion it is expected that the optimal ion backflow conditions will be fulfilled with a 1000 lpi mesh for 100 $\mu$m gap, and with a 1500 lpi mesh for 50 $\mu$m. In addition, it is expected that ion backflow will not be affected by a magnetic field, since ion masses make them insensitive to it.

\section{Measurements}  
\begin{figure}[h]
\centering{\resizebox{13.cm}{!}{\includegraphics{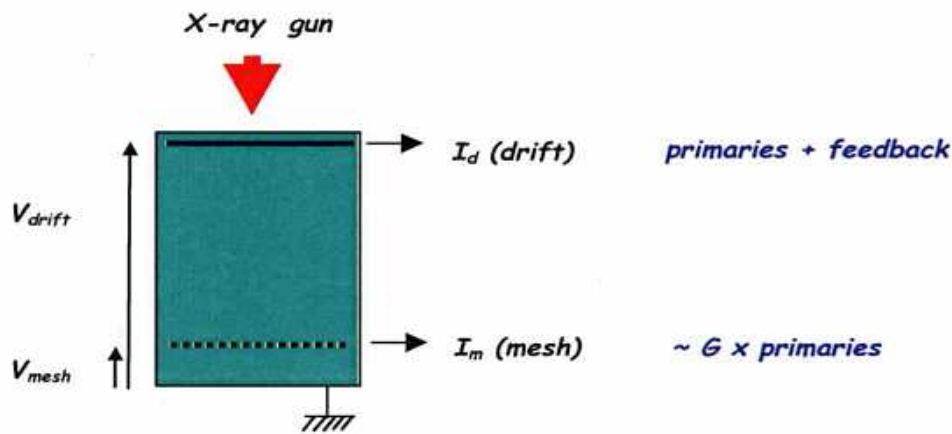}}}
\caption{Experimental device for ion feedback measurements}
\label{fig:Fig5}
\end{figure}

Measurements have been performed using an intense (10mA-10 keV) X-ray gun to produce primary electrons (see figure \ref{fig:Fig5}) in the 3mm drift space. The Ni micromesh, manufactured at CERN was located at a distance of 100 $\mu$m from the anode plane, and the typical gain was a few hundred. Gas mixture was Argon with 10\% isobutane or 2-3\% CH4.
Currents on the drift (i$_d$) and mesh (i$_m$) electrodes were accurately measured. The primary ionisation current i$_p$, which is of the order of a few 10pA, was obtained by measuring the drift current without gain (by lowering the voltage on the mesh). From these current measurements, it is easy to determine the ion backflow $\beta$: $\beta$ = (i$_d$-i$_p$)/(i$_d$+i$_p$) as a function of the field ratio by changing the voltage on the drift electrode ($\alpha$ was varying in a large dynamic range, between 10 and 700).

\vspace*{0.8cm}
 
\vspace*{0.6cm}

\begin{figure}[h]
\centering{\resizebox{11.cm}{!}{\includegraphics{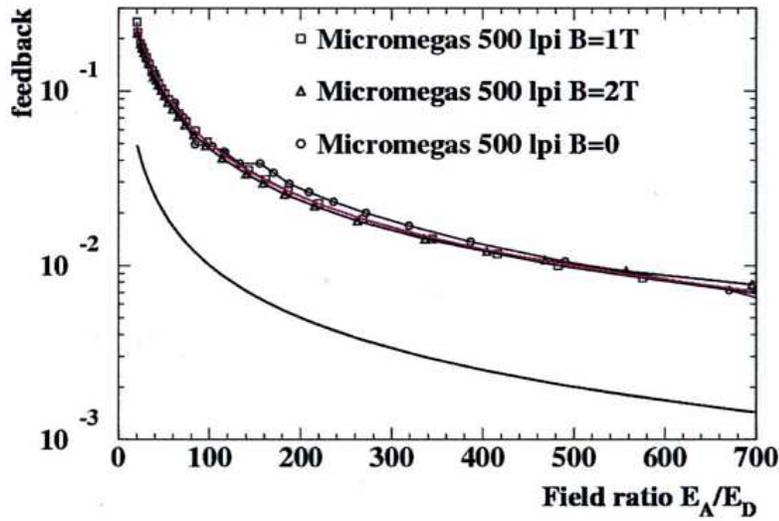}}}
\caption{Measurements of ion backflow vs field ratio for a 500 lpi micromesh}
\label{fig:Fig6}
\end{figure}

Figure \ref{fig:Fig6} shows measurements performed with a 500 lpi electroformed Ni mesh: as expected from calculations, the extension of the avalanche is not large enough compared to the mesh aperture, and ion backflow is degraded by a quite large factor ($\sim$4) as compared to 1/$\alpha$. 

\vspace*{0.8cm}
\vspace*{0.4cm}

\begin{figure}[h]
\centering{\resizebox{11.cm}{!}{\includegraphics{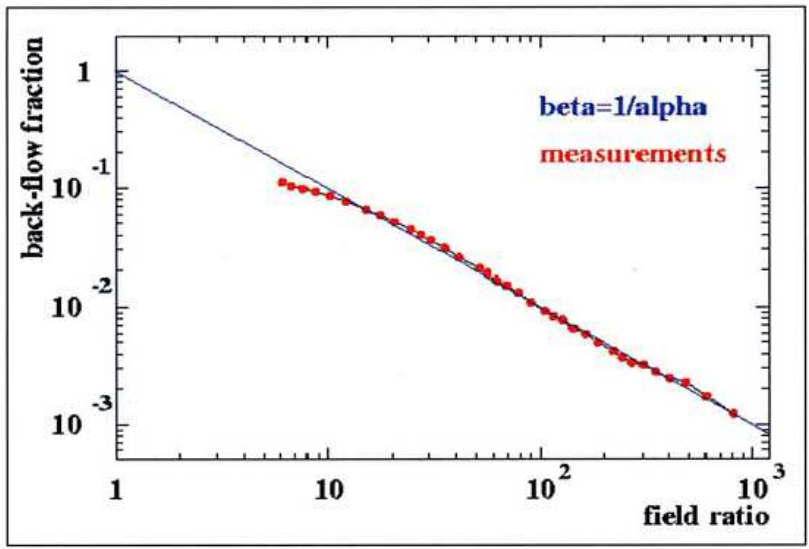}}}
\caption{Measurements of ion backflow vs field ratio for a 1500 lpi micromesh}
\vspace*{0.2cm}
\label{fig:Fig7}
\end{figure}

   Then measurements (see figure \ref{fig:Fig7}) have been done with a smaller pitch Ni mesh (1500 lpi, 17 $\mu$m): as expected, the backflow is exactly equal to the inverse of the field ratio over a very large range of field ratios.\\
\vspace*{0.8cm}
\vspace*{0.6cm}

\begin{figure}[h]
\centering{\resizebox{9.cm}{!}{\includegraphics{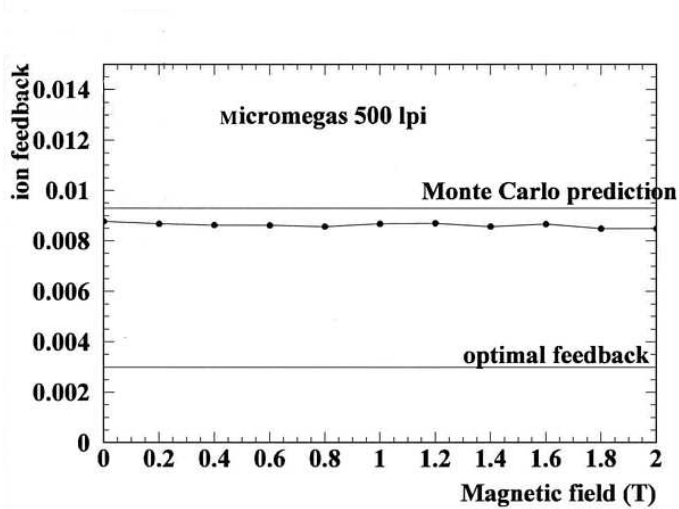}}}
\caption{Measurements of ion backflow vs magnetic field from 0 to 2T}
\label{fig:Fig8}
\end{figure}

   Finally, measurements have also been done in a superconducting coil, varying the magnetic field from 0 to 2T, without any change in the ion feedback, as expected (see figure \ref{fig:Fig8}).
\vspace*{0.8cm}

\section{Conclusion}    
   It has been proved and explained that in a Micromegas TPC device the ion backflow is equal to the inverse of the field ratio between the amplification and the drift electric fields, with only a few restrictions on the gas mixture, and on the mesh which should have a small pitch ($<25 \mu$m).
   As a conclusion, in the conditions of a future Micromegas TPC for the next linear collider, with a field ratio equal to 300-500, the expected ion feedback will be of the order of 2 $\times$ 10$^{-3}$. If the TPC can work at a relatively low gain ($<1000$), the total amount of secondary ions feeding the drift volume will be of the same order than the primary ionisation. Thus it is possible to envisage the construction of the detector without a gating grid, which is a major simplification in the design and the construction.

\section{Acknowledgements}
   We wish to thank J. Martin, J. Jeanjean and V. Puill for their contribution to this work.

\end{document}